\documentclass[3p,times,number]{elsarticle}

\usepackage{amssymb}

\usepackage{lineno}

\usepackage{amsmath}

\usepackage{float}
\usepackage{multirow}
\usepackage{booktabs} 
\usepackage{mathrsfs}

\usepackage{subcaption}

\usepackage[linesnumbered,vlined, ruled,commentsnumbered]{algorithm2e}
\usepackage{xcolor} 

\usepackage{qtree}
\usepackage{newfloat}
\DeclareFloatingEnvironment[fileext=lod]{diagram}

\begin{document}

\begin{frontmatter}

\title{Identifying combinations of tetrahedra into hexahedra: a vertex based strategy}

\author[UCL]{Jeanne Pellerin\corref{cor1}}
\ead{Jeanne.Pellerin@uclouvain.be}

\author[UCL]{Amaury Johnen}
\ead{Amaury.Johnen@uclouvain.be}

\author[UCL]{Kilian Verhetsel}
\ead{Kilian.Verhetsel@uclouvain.be}

\author[UCL]{Jean-Fran\c cois Remacle}
\ead{Jean-Francois.Remacle@uclouvain.be}

\address[UCL]{Universit\'e catholique de Louvain, iMMC, Avenue Georges Lemaitre 4, bte L4.05.02, 1348 
Louvain-la-Neuve, Belgium}

\cortext[cor1]{Corresponding author: Universit\'e catholique de Louvain, iMMC, Avenue Georges Lemaitre 4, bte L4.05.02, 1348 
Louvain-la-Neuve, Belgium. Tel.: +32 10 47 23 55. E-mail address: Jeanne.Pellerin@uclouvain.be}

\begin{abstract}
Indirect hex-dominant meshing methods rely on the detection of adjacent tetrahedra
that may be combined to form hexahedra, prisms and pyramids. 
In this paper we introduce an algorithm that performs this identification 
and builds the set $H$ of all possible combinations of tetrahedral elements of an input mesh $T$
into hexahedra, prisms, or pyramids. All identified cells are valid for engineering analysis.
First, all combinations of eight/six/five vertices whose connectivity
in $T$ matches the connectivity of a hexahedron/prism/pyramid are computed. 
The subset of tetrahedra of $T$ triangulating each potential cell is then determined.
Quality checks allow to early discard poor quality cells and to dramatically improve the efficiency of the method. 
Each potential hexahedron/prism/pyramid is computed only once.
Around 3 millions potential hexahedra are computed in 10 seconds on 
a laptop. We finally demonstrate that the set of potential hexes built by our algorithm is 
significantly larger than those built using predefined patterns of subdivision 
of a hexahedron in tetrahedral elements.

\end{abstract}


\end{frontmatter}

\section{Introduction}
 
\begin{figure}
	\centering
	\includegraphics[width=0.95\textwidth]{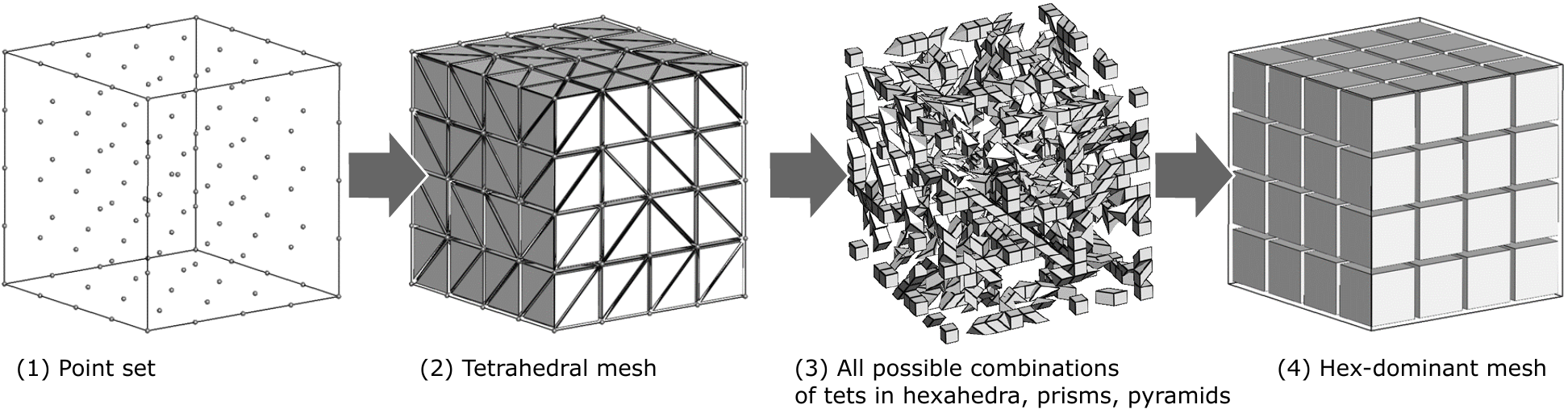}
	\caption{Indirect hexahedral dominant meshing principle.}
	\label{fig:indirect_meshing_workflow}
\end{figure}

In this paper we propose a new algorithm to identify all the hexahedra that may be built by combining tetrahedra of a given mesh.
Hexahedral meshes are considered by most of the finite element practitioners to
be superior to tetrahedral meshes (see e.g. \citep{benzley_comparison_1995}).
Yet, no robust meshing technique is able to process general 3D domains.
And generating hexahedral meshes in an automatic manner is still
considered as the ultimate goal in mesh generation \citep{shepherd_hexahedral_2008}.
Recently, promising techniques producing meshes composed of a majority of hexahedra
have been proposed
\citep{meshkat_generating_2000,  yamakawa_fully-automated_2003, baudouin_frontal_2014,
botella_indirect_2016, sokolov_hexahedral-dominant_2016}.
These methods take advantage of the existence of robust algorithms to generate
tetrahedral meshes and combine tetrahedra to produce meshes composed of a majority of hexahedra
associated to prisms, pyramids and tetrahedra .
The four steps of these indirect hex-dominant meshing methods can be summarized as follows (see also Figure~\ref{fig:indirect_meshing_workflow}):
\begin{enumerate} 
	\item A set of mesh vertices $V$ is initially sampled in the domain.
	\item A tetrahedral mesh $T$ is built by connecting $V$,  e.g. using a Delaunay kernel like \citep{si_tetgen_2015}.
	\item The set of potential cells $H$ (hexahedra, prisms, pyramids) that can be defined by 
		     combining tetrahedra of $T$ is built.
	\item A maximal subset $H_c \subset H$ constituted of cells that can be part of the same final mesh is determined.
	         The final \emph{hex-dominant mesh} is obtained adding the remaining not selected tetrahedra $T'$.	
\end{enumerate}
To reach the ultimate goal and combine all tetrahedra into hexahedra, i.e. obtain a final full-hexahedral mesh,
all steps are crucial. Previous works primarily focus on the first step of placing the
final mesh vertices.
In this paper, we focus on the third step. Our input is a tetrahedral mesh $T$ of a given point set $V$ 
and we output the set of all hexahedra, prisms, and pyramids that may be built by combining 
tetrahedra of $T$.

\begin{figure}
	\centering
	\includegraphics[width=0.5\textwidth]{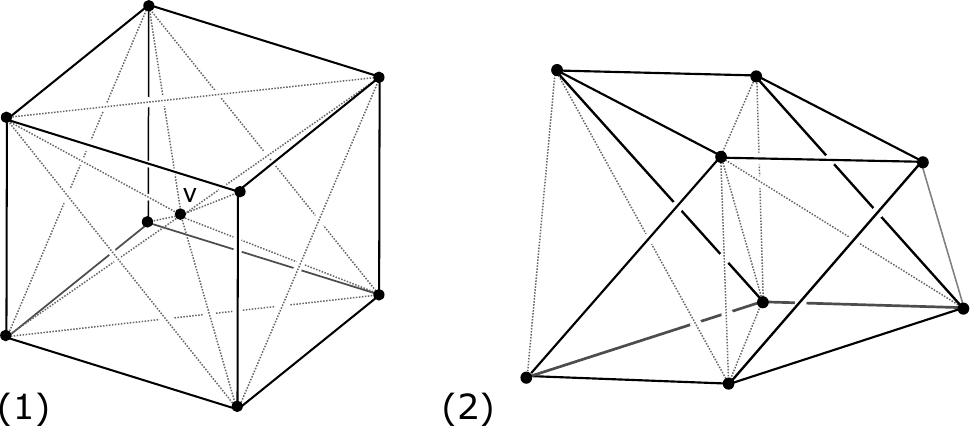}
	\caption{Two hexahedra not identified by existing combination methods.
	(1) A decomposition with an interior vertex v. 
	(2) A decomposition into eight tetrahedra. This is a counter example to \citep{sokolov_hexahedral-dominant_2016}'s 
	claim that there is no decomposition of the hexahedron into more than seven interior tetrahedra.
	}
	\label{fig:motivation_ignored_hexes}
\end{figure}

Identifying combinations of tetrahedra into pyramids is trivial. There are three possible pyramids for each 
facet shared by two tetrahedra of $T$. Identifying combinations of tetrahedra 
into prisms is a bit more challenging since three tetrahedra properly connected should be identified. 
However, when identifying combinations of tetrahedra into hexahedra, 
there are at least ten different subdivisions of a hexahedron into
five, six, or seven tetrahedra (\S \ref{sec:background}). To overcome this challenge, 
two main approaches have been proposed.
The first relies on a predefined set of patterns of the decomposition of a hexahedron into tetrahedra 
\citep{meshkat_generating_2000, botella_indirect_2016, sokolov_hexahedral-dominant_2016},
the second on patterns of edge connections in a hexahedron \citep{yamakawa_fully-automated_2003, baudouin_frontal_2014}.
Their main limitation is that they do not build the largest set of potential hexahedra $H$.
 For example, they do not detect the two hexahedra of Figure~\ref{fig:motivation_ignored_hexes} (more details in \S\ref{sec:real_motivations}).

In this paper, we introduce an algorithm that detects all possible combinations of tetrahedra into hexahedra.
The algorithm is based on the local search of combinations of eight vertices 
that are adequately connected to build a hexahedron.
The key advantages of the new algorithm are that it computes all possible potential hexahedra,
computes each of them once only, discards bad quality hexes at an early stage, 
is easy to implement and is very efficient.
The algorithm does not rely on any pattern.
An algorithm variation permits to compute all the possible potential prisms.

After reviewing the main methods to combine tetrahedra into hexahedra , prisms or pyramids
(\S\ref{sec:background}), 
we detail our vertex-based algorithm to identify the potential hexahedra , prisms and pyramids 
in a tetrahedral mesh  (\S\ref{sec:the_algo}). 
We demonstrate we actually compute the set of all possible hexahedra , prisms, and pyramids.
We further show in particular that the set of hexahedra built by our algorithm 
is larger than the one built by existing methods (\S\ref{sec:comparison}).
Examples of hex-dominant mesh that may be generated from these potential cells are given 
in \S\ref{sec:hex_dominant_meshes}.
The C++ code implementing the methods of this paper is open-source and available at https://www.hextreme.eu/download/.

\section{Background}
\label{sec:background}

Before giving details on subdivisions of hexahedra , prisms or pyramids into tetrahedra (\S\ref{sec:hex_decompositions_background}) 
and on methods used to identify these in an existing mesh (\S\ref{sec:state_of_the_art}), 
we define the terms and notations used throughout the paper.
  
\subsection{Definitions}
\label{sec:definitions}

\begin{figure}
	\centering
	\includegraphics[width=0.75\textwidth]{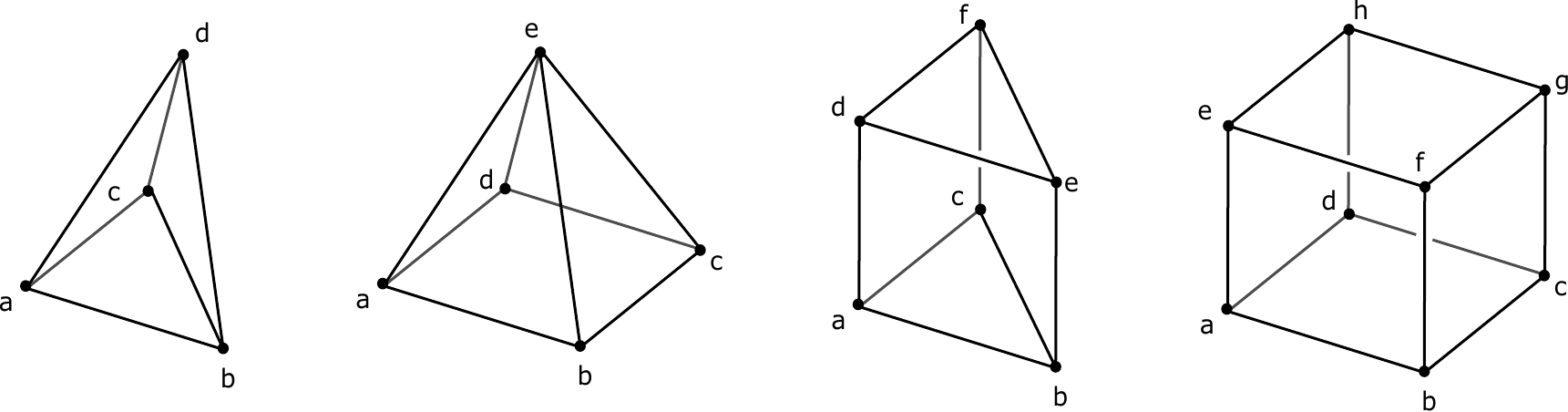}
	\caption{3D cell templates: tetrahedron, pyramid, prism, and hexahedron.}
	\label{fig:templates}
\end{figure}

We have to be very clear that all the cells we are considering are
finite element cells hexahedra /prisms/pyramids valid for finite element simulations.
A very important point is that their quadrilateral facets are not planar, but are bilinear surfaces.
We require the Jacobian determinant to be strictly positive at any point inside the cell.
The following conventions will be used throughout this paper (see Figure~\ref{fig:templates}).
\newline
%
The \emph{pyramid} \{abcde\} has: 
\begin{itemize}
	\item 5 vertices: \{a\}, \{b\}, \{c\}, \{d\}, \{e\}
	\item 8 straight line edges: 
			\{ab\}, \{bc\}, \{cd\}, \{ad\}, 
			\{ae\}, \{be\}, \{ce\}, \{de\}
	\item 4 planar triangle faces: \{abe\}, \{bce\}, \{cde\}, \{ade\}
	\item 1 bilinear quadrilateral face:	\{abcd\}
\end{itemize}
The \emph{prism} \{abcdef\} has: 
\begin{itemize}
	\item 6 vertices: \{a\}, \{b\}, \{c\}, \{d\}, \{e\}, \{f\}
	\item 9 straight line edges: 
			\{ab\}, \{bc\}, \{ca\},
			\{ad\}, \{be\}, \{cf\}, 
			\{de\}, \{ef\}, \{fd\} 
	\item 2 planar triangle faces: \{abc\}, \{def\}
	\item 3 bilinear quadrilateral faces:	\{abed\}, \{efcb\}, \{acfd\}
\end{itemize}
The \emph{hexahedron} \{abcdefgh\} has: 
\begin{itemize}
	\item 8 vertices: \{a\}, \{b\}, \{c\}, \{d\}, \{e\}, \{f\}, \{g\}, \{h\}, 
	\item 12 straight line edges: 
			\{ab\}, \{bc\}, \{cd\}, \{ad\}, 
			\{ae\}, \{bf\}, \{cg\}, \{dh\}, 
			\{ef\}, \{fg\}, \{gh\}, \{eh\} 
	\item 6 bilinear quadrilateral faces:	\{abcd\}, \{efgh\}, \{abfe\}, \{dcgh\}, \{bcgf\}, \{adhe\},
\end{itemize}

\begin{figure}
	\centering
	\includegraphics[width=0.45\textwidth]{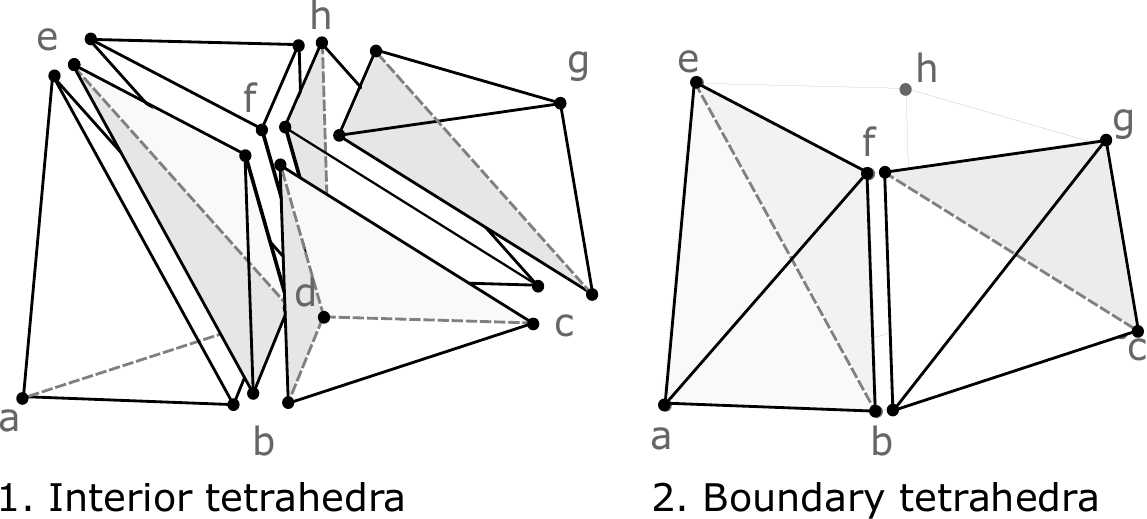}
	\caption{
		Interior tetrahedra and boundary tetrahedra of a hexahedron triangulation. 
		Boundary tetrahedra connect the four vertices of the same quadrilateral facet.
	}
	\label{fig:definitions}
\end{figure}

A \emph{triangulation} (tetrahedrization) of a hexahedron/prism/pyramid is a triangulation 
of the vertices of the cell that respect the cell boundary,  in other words 
it is a subdivision of the interior of the hexahedron/prism/pyramid
into a set of conformal tetrahedra without any additional vertex. 
The tetrahedra induce a subdivision of each quadrilateral facet into two triangles
by a diagonal boundary edge.
We further define the \emph{boundary tetrahedra} as
the tetrahedra connecting the four vertices of a cell quadrilateral facet (Figure~\ref{fig:definitions}.2). 
In previous works \citep{meshkat_generating_2000, botella_indirect_2016, baudouin_frontal_2014, 
sokolov_hexahedral-dominant_2016},
boundary tetrahedra are called slivers. We do not use that term which refers 
to a geometrical property (degeneracy) of tetrahedra.
The triangulation is determined by \emph{interior tetrahedra} (Figure~\ref{fig:definitions}.1).
Indeed, the addition or removal of one or several boundary tetrahedra  does not modify the cell.

\subsection{Decomposing a cell into tetrahedral elements}
\label{sec:hex_decompositions_background}

We first review the subdivisions of the pyramid, of the prism and of the 3-cube into tetrahedra 
\citep{de_loera_triangulations:_2010}.

\paragraph{Triangulation of the pyramid} There are as many subdivisions of the general pyramid as there are 
subdivisions of its planar base \citep{de_loera_triangulations:_2010}.   
We are considering square pyramids, there are then exactly two triangulations of the pyramid.

\paragraph{Triangulation of the prism} The ordinary  triangular prism is the result of the product 
of a triangle with an edge: prism$(D_3) = D_3 \times D_2$. 
A prism($D_n$) has exactly $n!$ triangulations that are all equivalent to one another by affine symmetries \citep{de_loera_triangulations:_2010}. 
The prism has then 6 triangulations that are all equivalent.

\begin{figure}
	\centering
	\includegraphics[width=0.9\textwidth]{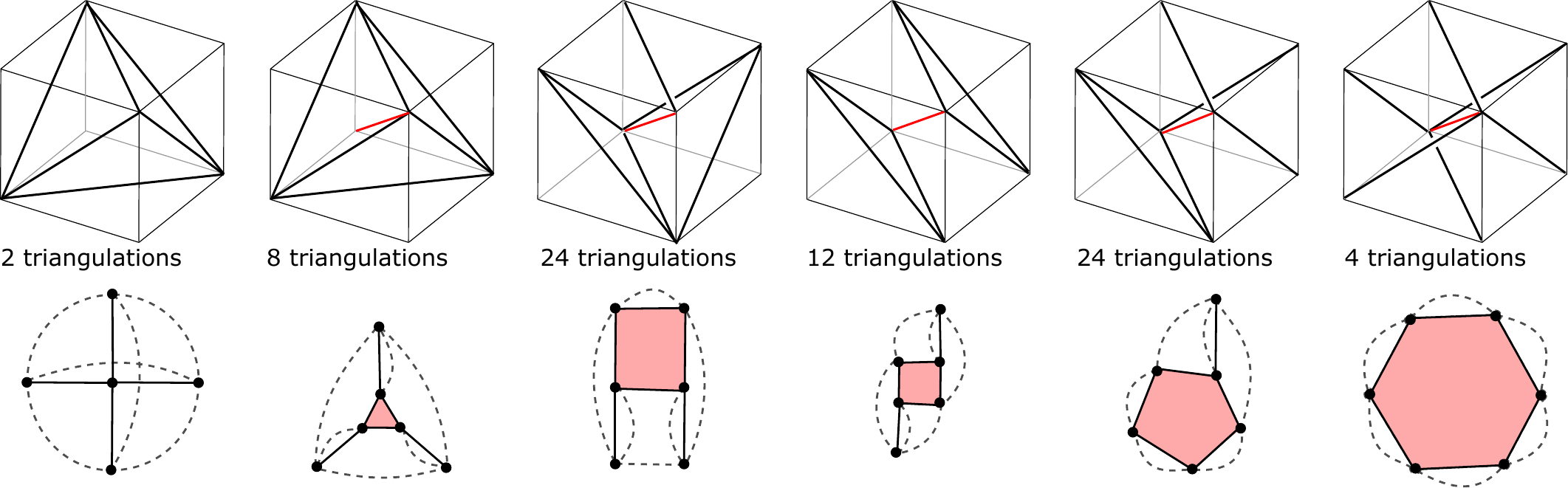}
	\caption{ 
		The six types of triangulations of the 3-cube and their dual complex representations.
		In the dual complexes, there is on vertex per tetrahedron , one plain edge linking adjacent tetrahedra 
		and one dashed edge linking tetrahedra incident to the same quadrilateral facet.
		The interior (red) edge is the cube diameter and corresponds to a cell in the dual complex.
	}
	\label{fig:perfect_cube_decompositions}
\end{figure}

\paragraph{Triangulation of the 3-cube I$^3 = {[0,1]^{3}}$} 
The 3-cube has exactly 74 triangulations \citep{de_loera_triangulations:_2010} : 
\begin{enumerate}
	\item Every triangulation of the 3-cube contains either a regular tetrahedron 
    (i.e. a tetrahedron whose 6 edges are of equal lengths) or a diameter, i.e. an interior edge joining two
	opposite vertices (red edges on Figure~\ref{fig:perfect_cube_decompositions}).
	\item There are 2 triangulations with a regular tetrahedron, symmetric to one another.
        The triangulations containing an interior edge are completely classified modulo symmetries by
        their dual complex which can be one of the last five shown on Figure~\ref{fig:perfect_cube_decompositions}.
        There are respectively 8, 24, 12, 24, 4 triangulations in each class.
\end{enumerate}
A dual complex (Figure~\ref{fig:perfect_cube_decompositions}) is a practical way to visualize 
the 6 different possible decompositions (tetrahedrizations) of the 3-cube. 
In the dual  complex, also called dual graph, one vertex corresponds to one tetrahedron
and two vertices are connected by an edge if the 
corresponding tetrahedra are adjacent through a triangular facet. A 2-cell of the dual complex 
(cycle in the dual graph) corresponds to an interior edge of the tetrahedrization (red on Figure~\ref{fig:perfect_cube_decompositions}).
In a meshing context, these different possible decompositions
of the 3-cube were identified by \cite{meshkat_generating_2000} 
who enumerate the feasible dual complex graphs, called RF-graph in their paper.
In the RF-graph, additional dashed edges connect tetrahedra that 
are adjacent to the same quadrilateral facet (Figure~\ref{fig:perfect_cube_decompositions}).

\begin{figure}
	\centering
	\includegraphics[width=0.6\textwidth]{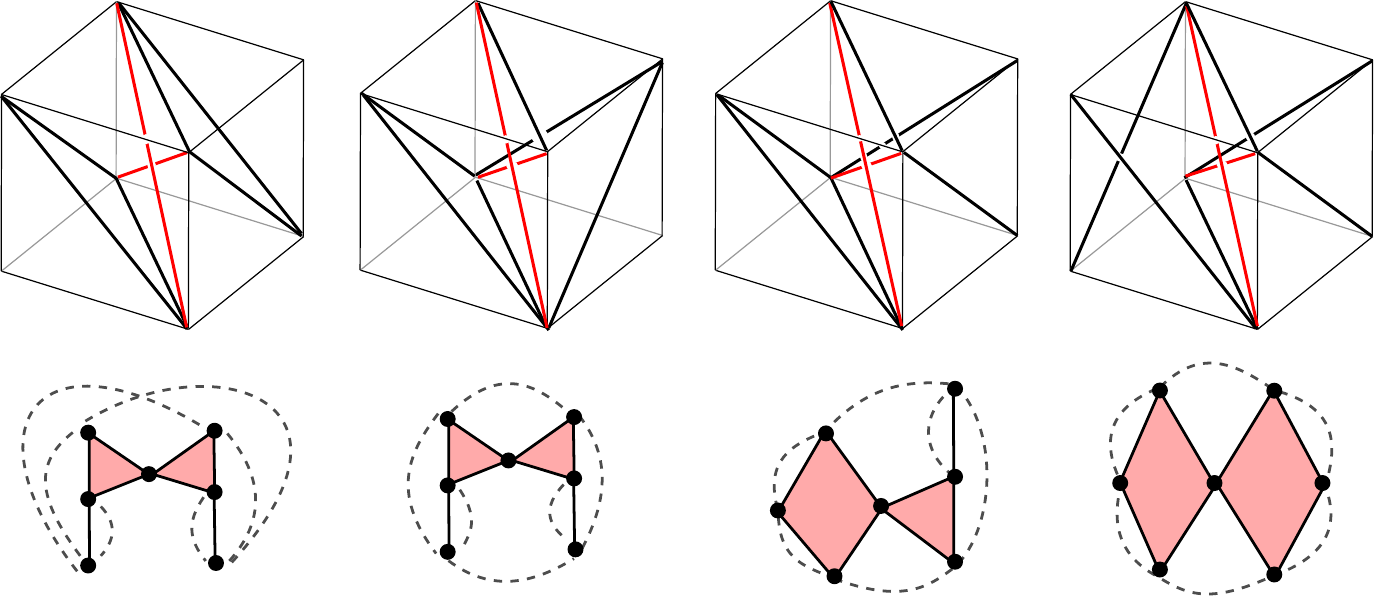}
	\caption{
		The four types of triangulations of an almost perfect cube into 7 tetrahedra proposed by \citep{botella_indirect_2016}
		and their dual complex representation.		
	}
	\label{fig:7_tet_decompositions}
\end{figure}

\paragraph{Triangulations of the real cube}
Recently, the work of \cite{meshkat_generating_2000} was extended by \cite{botella_indirect_2016} 
and \cite{sokolov_hexahedral-dominant_2016} who proposed four additional decomposition patterns  
into seven tetrahedra (Figure~\ref{fig:7_tet_decompositions}).
The hexahedron is split into two prisms by a tetrahedron without any facet on the hexahedron boundary 
and containing two interior edges. For this tetrahedron to have a strictly positive volume, it is sufficient 
to work in finite precision, i.e. move slightly one of its vertices. 

\paragraph{Bounds on the number of tetrahedra}
The Euler characteristic gives a relationship between the number of tetrahedra~$t$ 
in a cell decomposition and its number of interior edges $e_{int}$. 
Each triangulation is a 3-ball with Euler characteristic $\chi=1$,
where $\chi = v - e + f - t$.
Then $ v - e_{int} - e_{bd} + f_{int} +f_{bd} - t= 1$.
Since there are 4 triangular faces per tetrahedron and 2 tetrahedra per interior triangular face, 
we have $4t = 2 f_{int} + f_{bd}$.
Since the number of boundary edges $e_{bd}$, and boundary triangular facets $f_{bd}$ are fixed for each type of cell, 
the number of tetrahedra $t$ in a hexahedron , prism or pyramid decomposition 
(without internal vertices) depends only on the number of interior edges $e_{int}$.
Moreover, there are at most ${{n}\choose{2}} - e_{bd}$ edges in a cell with $n$ vertices
we then have trivial bounds on the number of tetrahedra in the triangulation of the cells.
\begin{itemize}
 \item For the hexahedron: $t_{hex} = 5 +e_{int}$ and  $5 \leq t_{hex} \leq 15$.  
 \item For the prism: $t_{prism} = 3 +e_{int}$ and  $3 \leq t_{prism} \leq 6$ 
 \item For the pyramid: $t_{prism} = 2 +e_{int}$ and  $2  \leq t_{pyramid} \leq 3$. 
\end{itemize}
See also \cite{edelsbrunner_tetrahedrizing_1990} for additional combinatorial results.

\subsection{Combining tetrahedra into hexahedra: state of the art}
\label{sec:state_of_the_art}

To compute the set $H$ of potential hexahedra and other cells that may be built by combining the 
elements of a tetrahedral mesh $T$ without modifying its connectivity there are two known approaches. 
\cite{meshkat_generating_2000} propose to find combinations of tetrahedra 
into hexahedra by searching the adjacency graph of $T$ for all occurrences of 
the cube decomposition dual complexes (Figure~\ref{fig:perfect_cube_decompositions}).
The problem of matching subgraphs in large sparse graphs is solved
using standard data mining algorithms that operate on graphs.
The same technique is used by \cite{levy_l_2010, huang_boundary_2011, botella_indirect_2016} 
and \cite{sokolov_hexahedral-dominant_2016} 
who consider four decompositions into seven tetrahedra (Figure~\ref{fig:7_tet_decompositions}). 
The second approach proposed by \cite{yamakawa_fully-automated_2003}
relies on the vertices and edges of the tetrahedral mesh $T$.
Local searches are performed into the vertex-edge graph of $T$ using two patterns.
These vertex connectivity patterns generalize those proposed by \cite{meshkat_generating_2000}
and relax partially the dependency on the tetrahedral mesh.
This method has been implemented by \cite{baudouin_frontal_2014} where a 
third pattern taking into account configurations with an interior flat tetrahedron was added.
Some other approaches like H-Morph \citep{owen_h-morph:_2000} 
combine tetrahedra into hexahedra, while allowing for modifications
of the connectivity and geometry of the input tetrahedral mesh (tetrahedron flips, node insertions, and 
node displacement).  This great flexibility can make the algorithm intractable, but 
one advantage is that it maintains a valid mixed mesh throughout the procedure.

\subsection{Motivations for a new approach}
\label{sec:real_motivations}

Important observations led us to work on improving these existing
techniques.
First, they do not identify the largest set $H$ of potential hexahedra. 
On Figure~\ref{fig:motivation_ignored_hexes} we gave two valid hexahedra 
that would neither be found by \cite{meshkat_generating_2000}'s method
nor by \cite{yamakawa_fully-automated_2003}'s method. 
The first is a decomposition that encompasses one internal vertex, a configuration 
that may occur when a Steiner point is added when generating the tetrahedra.
The second is a decomposition that has 8 interior tetrahedra.
It is a counter example to \cite{sokolov_hexahedral-dominant_2016}'s claim that 
there is no hexahedron decomposition with more than 7 interior tetrahedra.
Both decompositions are not identified when searching 
for hexahedra made of 5, 6, or 7 interior tetrahedra.
Neither are they by \cite{yamakawa_fully-automated_2003}'s algorithm
since none of their constitutive tetrahedra has three facets on the hexahedron boundary.

\begin{figure}
	\centering
	\includegraphics[width=0.4\textwidth]{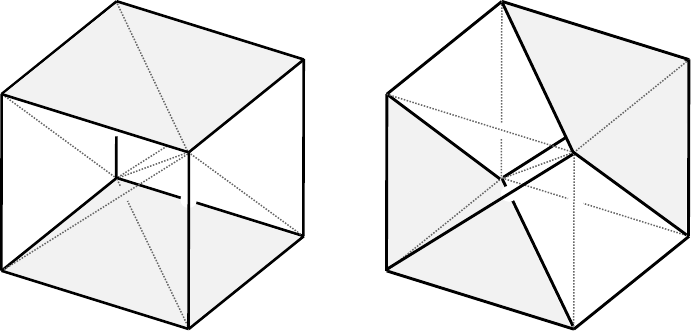}
	\caption{
		Two hexahedra with different edges and faces may be defined from the same set of tetrahedra.
	}
	\label{fig:two_hex_one_pattern}
\end{figure}

Second, as mentioned by \cite{yamakawa_fully-automated_2003}, 
 several hexahedra may be defined using the same decomposition
pattern by modifying the ordering of the vertices (Figure~\ref{fig:two_hex_one_pattern}). 
The hexahedra have different edges and different faces while having the same tetrahedral decomposition.
The hexahedron on the left being a perfect cube, the one on the right is undoubtedly invalid
(zero Jacobian determinant), but were the vertices in a more general position,
both could be valid.
Third, the existing methods  identify the same hexahedron several times.
That number is as high as the number of corner tetrahedra in the decomposition in
\cite{yamakawa_fully-automated_2003}'s approach and depends on dual complex 
symmetries in \cite{meshkat_generating_2000}'s approach.
With those observations in mind, we believe that an algorithm that
finds all potential hexahedra in a tetrahedral mesh should not be based
on a predefined set of patterns.

\section{An algorithm to combine tetrahedra into hexahedra}
\label{sec:the_algo}

In this section, we detail our algorithm to detect combinations of tetrahedra into hexahedra or prisms 
and  find all cells that may be generated by combining elements of a given input tetrahedral mesh $T$.
We first explain the 2d version of the algorithm  that combines triangles into quadrilaterals 
and show its relationship to algorithms generating permutations and combinations.

\subsection{The 2D algorithm}
\label{sec:2d_algo}

The 2D version of the algorithm
is built by modifying an algorithm generating the $4$-permutations of a $n$-set.
Given a set $V$ of  $n$ vertices labeled from $1$ to $n$, let us first 
 compute all possible quadrilaterals that can be defined from these vertices. 
We define a numbered quadrilateral $abcd$ by the order of its 4 vertices and we define
a non-oriented quadrilateral $abcd$ by its edges $ab, bc, cd$ and $da$ ignoring orientation.
Generating all numbered quadrilaterals that can be built from $V$ is a combinatorial problem 
solved by  Algorithm~\ref{algo:stupidest_2D_algorithm} which generates all possible permutations
of 4 vertices of the $n$ labels. 
When $V=\{1,2,3,4\}$, the output is the set of the 24 permutations of 4 values (Table~\ref{tab:24_permutations}).
These 24 permutations define 3 different non-oriented quadrilaterals, each of them corresponding to 8 equivalent permutations.
Adding constraints on the relative order of the vertices of one quadrilateral 
$a<b$, $a < c$, $b<d$, we obtain Algorithm~\ref{algo:less_stupid_2D_algorithm} 
which fulfill our first objective and compute all possible non-oriented quadrilaterals that can be built from $V$.
The output for $V=\{1,2,3,4\}$ is now the 3 non-oriented quadrilaterals that may be generated from 4 vertices 
(Table~\ref{tab:the_3_quads}).

\vspace{0.25cm}
 \begin{minipage}{.45\linewidth}
		{ 
			\begin{algorithm}[H]
			\small   
			\SetKwData{Quad}{quad}
			
			\KwData{$V$ vertex set}
			\KwResult{$Q$ set of potential quads}
			\BlankLine
			\ForEach{$a$ in $V$}{                                             		
				\ForEach{$b$ in $V$, $b \neq a$ }{                       
					\ForEach{$c$ in $V$, $c \notin \{a,b\}$}{     
						\ForEach{$d$ in $V$, $d \notin \{a, b, c\}$}{ 	
							$Q$ $\leftarrow$ $Q$ $\cup$ $\{a,b,c,d\}$\;
						}			
					}
				}
			}	
			
		\caption{4-permutations in $V$.}		
		\label{algo:stupidest_2D_algorithm}
		\end{algorithm} \par
		}
\end{minipage}
\hfill
 \begin{minipage}{.45\linewidth}
	{
		\begin{algorithm}[H]
		\small
		\SetKwData{Quad}{quad}
		
		\KwData{$V$ vertex set}
		\KwResult{$Q$ set of potential quads}
		\BlankLine
		\ForEach{$a$ in $V$}{                                             
			\ForEach{$b$ in $V$, $b >a$ }{                       
				\ForEach{$c$ in $V$,  $c >a$, $c \neq b$}{     
					\ForEach{$d$ in $V$,  $d >b$, $d \neq c$ }{ 	
						$Q$ $\leftarrow$ $Q$ $\cup$ $\{a,b,c,d\}$\;
					}			
				}
			}
		} 
		\caption{Unique quadrilaterals in $V$.}		
		\label{algo:less_stupid_2D_algorithm}
		\end{algorithm}
	}
\end{minipage}

 \begin{minipage}{.45\linewidth}
	\begin{table}[H]
		\centering
		\begin{tabular}{cccccc}
			\color{gray} 1234 & \color{black} 1243 & \color{purple} 1324 & \color{black} 1342 & \color{purple} 1423 & \color{gray} 1432 \\
			\color{black} 2134 & \color{gray} 2143 & \color{purple} 2314 & \color{gray} 2341 & \color{purple} 2413 & \color{black} 2431 \\
			\color{black} 3124 & \color{purple} 3142 & \color{gray} 3214 & \color{purple} 3241 & \color{gray} 3412 & \color{black} 3421 \\
			\color{gray} 4123 & \color{purple} 4132 & \color{black} 4213 & \color{purple} 4231 & \color{black} 4312 & \color{gray} 4321 \\
		\end{tabular}
		\caption{Output of Algorithm~\ref{algo:stupidest_2D_algorithm} for $V=\{1,2,3,4\}$ is the set of the 
					24 permutations of 4 values define 3 quadrilaterals.}
		\label{tab:24_permutations}%
	\end{table}
	\vspace{0.3cm}
\end{minipage}
\hfill
 \begin{minipage}{.45\linewidth}
 \begin{table}[H]
		\centering
		\begin{tabular}{ccc}
			\color{gray} 1234 & \color{black} 1243 & \color{purple} 1324 
			\vspace{1cm}
		\end{tabular}
		\caption{Output of Algorithm~\ref{algo:less_stupid_2D_algorithm} for $V=\{1,2,3,4\}$ is the set
		of the 3 possible quadrilaterals.}
		\label{tab:the_3_quads}%
	\end{table}
	\vspace{0.3cm}
\end{minipage}

\begin{figure}
	\centering
	\includegraphics[width=0.65\textwidth]{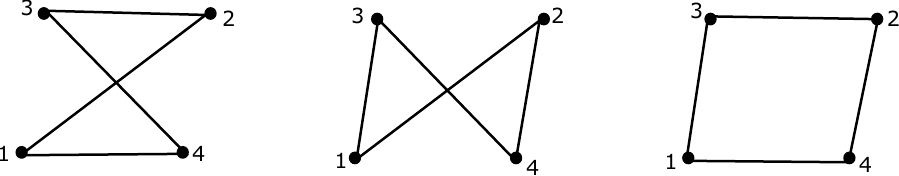}
	\caption{For 4 points in $\mathbb{R}^2$, only 1 over 3 possible combinatorial quadrilateral is valid.
	Note that in  $\mathbb{R}^3$, all 3 quadrilaterals define valid bilinear quadrilateral facets.
	}
	\label{fig:2D_quads}
\end{figure}

\begin{figure}
\begin{subfigure}[b]{0.40\textwidth}
	\centering
	\includegraphics[width=0.75\textwidth]{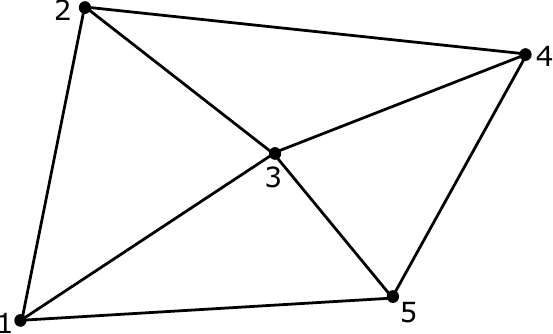}
	\caption{}	
	\label{fig:2D_example}
\end{subfigure}
\begin{subfigure}[b]{0.45\textwidth}
\footnotesize
	\centering
		\Tree 
		[.$\emptyset$	
			[.1 
				[.2 [.3 4 ] [.5 3 4 ]]
				[.3 [.5 4 ] ]
				5 
			] 
			[.2 
				[.3 [.4 5 ] ]
				4
			]
			[.3 [.4 5 ] 5 ]
			[.4 5 ]
			5 
		] 	
	\caption{}
	\label{fig:2D_tree}
	\end{subfigure}
	\caption{Search tree of Algorithm~\ref{algo:2D_algorithm} on a example. Left: Input 2D mesh. 	
				Right: Each branch reaching a depth of 4 defines a quadrilateral. 
				Five quadrilaterals are identified: \{1243\}, \{1235\}, \{1345\}, \{1245\}, \{2354\}.}
\end{figure}

Let us now associate to each labeled vertex a point of $\mathbb{R}^2$.
Among the 3 possible quadrilaterals that can be defined from 4 points of $\mathbb{R}^2$, only 
one is valid, i.e. has non-intersecting edges (Figure~\ref{fig:2D_quads}).
We further associate to the set of vertices $V$ a triangulation $T$ and 
modify Algorithm~\ref{algo:less_stupid_2D_algorithm} such that it generates
all quadrilaterals whose edges are edges of the triangulation (Algorithm~\ref{algo:2D_algorithm}).
The search for $b$ and $d$ is then restricted to 
the set of vertices connected to $a$. Similarly $c$ should be 
connected through an edge to both $b$ and $d$.
The last step of the procedure is to identify the triangles subdividing each quadrilateral.
Vertex selection order is now $a,b,d,c$ instead of $a,b,c,d$ since the choice of
$c$ depends on both $b$ and $d$.
All steps of the identification of quadrilaterals in a simple mesh are detailed 
on Figure~\ref{fig:2D_tree}.

The advantage of this approach over the classical algorithms pairing adjacent triangles 
(e.g. \cite{remacle_blossom-quad:_2012}) is that it identifies quadrilaterals which encompass one (or more) vertex.
An example is the quadrilateral $\{1245\}$ on Figure~\ref{fig:2D_example}  that encompasses vertex~$\{3\}$.
The other advantages of the algorithm are that it is easy to add geometrical quality tests 
(edge lengths, angles of the quadrilateral under construction) and that 
its parallelization is trivial.
Its complexity may seem prohibitive but a vertex of a 2d triangulation is 
connected to an average of 6 other vertices.


\begin{algorithm}[h]
	\small
\SetKwData{Quad}{quad}
\SetKwFunction{Neighbors}{neighbors}
\SetKwFunction{HasEdge}{hasEdge}

	\KwData{$T$ triangulation of vertex set $V$}
	\KwResult{$Q$ set of potential quads}
	\BlankLine
	\ForEach{$a$ in $V$}{  
		\ForEach{$b$ in {\Neighbors{$a$}}, $b >a$ } {                                       
			\ForEach{$d$ in {\Neighbors{$a$}}, $d > b$ } {                             
				\ForEach{$c$ in {\Neighbors{$\{b,d\}$}}, $c > a$ } {        
					$Q$ $\leftarrow$ $Q$ $\cup$ $\{abcd\}$
				}			
			}
		}
	}
	\caption{Vertex based search algorithm of all potential quadrilaterals in a triangulation.}		
	\label{algo:2D_algorithm}
\end{algorithm}


\subsection{The 3D algorithm to combine tetrahedra into hexahedra}
\label{sec:3d_algo}

In this section we extend the 2D algorithm in 3D where  the goal is to identify combinations of tetrahedra into hexahedra.
Similarly to what we did in 2D, we can modify an algorithm generating all possible $8$-subset of 
the set of labeled vertices $V$ to generate exactly once all oriented hexahedra (Algorithm~\ref{algo:less_stupid_3D_algorithm}).
The first corner of the hexahedron is built by choosing vertices \{a, b, d, e\} such that $b>a$,  $d>b$ and $e>b$.
This corner sets the orientation of the hexahedron (Figure~\ref{fig:templates}). 
Orientation can be ignored by setting $e > d$.
The four other vertices are chosen to be greater than $a$.
The output of Algorithm~\ref{algo:less_stupid_3D_algorithm} for
$V=\{1,2,3,4,5,6,7,8\}$ is a set of 1,680 oriented hexahedra.
This is consistent with the well-known fact that there are 24 permutations of 
the labeled vertices that do not modify the orientation, the edges, or the faces of a hexahedron since 
$8! = 1,680 \times 24$.
When orientation is ignored, there are $48$ such permutations and then 840 different hexahedra.

The complexity of Algorithm~\ref{algo:less_stupid_3D_algorithm} is catastrophic, but it 
may be nonetheless useful for relatively small point sets.
For example, we managed to recompute the full-hex mesh of the 
Schneider's pyramid subdivision of \cite{yamakawa_88-element_2010} 
from its 104 points taking into account strong quality constraints 
on the hexahedra generated.

\begin{algorithm}[t]
\small
	\SetKwData{Hexes}{hexes}

	\KwData{$V$ vertex set}
	\KwResult{$H$ set of potential quads}
	\BlankLine
	\ForEach{$a$ in $V$}{                                             		
		\ForEach{$b$ in $V$, $b > a$ }{                       
			\ForEach{$d$ in $V$, $d > b$}{     
				\ForEach{$e$ in $V$, $e > b$}{ 	
					\ForEach{$c$ in $V$, $ c > a, e \notin \{b, d, e\}$} {                                             		
						\ForEach{$f$ in $V$, $f > a, f \notin \{b, d, e, c\}$}{                       
							\ForEach{$h$ in $V$, $h > a, h \notin \{b, d, e, c, f\}$}{                       
								\ForEach{$g$ in $V$, $g >a, g \notin \{ b, d, e, c, f, h\}$}{                       
									$H$ $\leftarrow$ $H$ $\cup$ $\{abcdefgh\}$\;
								}
							}
						}
					}
				}			
			}
		}
	} 
	\caption{Generates exactly once each potential oriented hexahedron in a vertex set $V$.}		
	\label{algo:less_stupid_3D_algorithm}
\end{algorithm}

\paragraph{Our algorithm} Let us now associate to each labeled vertex of $V$ a point of $\mathbb{R}^3$.
Among the ${{n}\choose{8}}\times1680$ possible hexahedra, only a small proportion 
will be geometrically valid. 
As we did in 2D with Algorithm~\ref{algo:2D_algorithm}, we restrict the search space 
using a triangulation (tetrahedrization) of the vertices  
and consider only hexahedra which edges are edges of the triangulation.
The final Algorithm~\ref{algo:3D_algorithm} outputs all the hexahedra that may be built using the edges 
of the triangulation.
Each hexahedron  is generated once only since it is a direct modification of Algorithm~\ref{algo:less_stupid_3D_algorithm}.
Moreover, by construction, Algorithm~\ref{algo:3D_algorithm} computes all the hexahedra that can be generated by 
combining tetrahedra of an input mesh~$T$.

The algorithm to identify prisms is built with the same principles. 
To build pyramids it is however much faster to iterate on all triangular facets of $T$
and to create three pyramids for each of them by changing the apex to be one of the three 
vertices of the facet.

\begin{algorithm}[t]
	\small
   \SetKwData{HexTets}{tets\_in\_hex}
   \SetKwFunction{Neighbors}{neighbors}
   \SetKwFunction{IsQuadFace}{is\_quad\_face}
   \SetKwFunction{ComputeTets}{compute\_tets}
   \SetKwFunction{CheckTwo}{check\_quad\_corner\_quality}
   \SetKwFunction{CheckThree}{check\_hex\_corner\_quality}
   
   	\KwData{$V$ vertex set, $T$ tetrahedrization of $V$}
	\KwResult{$H$ set of potential hexahedra}
	\BlankLine
	\ForEach{$a$ in $V$}{                                                      
		\ForEach{$b$ in \Neighbors{$a$}, $b>a$}{                       
			\ForEach{$d$ in \Neighbors{$a$}, $d>b$}{                        
				%
				\ForEach{$e$ in \Neighbors{$a$}, $e>b, e \neq d$}{         
					%
					\ForEach{$c$ in \Neighbors{$\{b,d\}$}, $c>a, c \neq e$}{      
						\lIf{ $!$\IsQuadFace{$a,b,c,d$}} {continue}												
						\ForEach{$f$ in \Neighbors{$\{b,e\}$}, $f >a, f \notin \{b,d,e,c\}$}{  
							\lIf {$!$\IsQuadFace{$a, b, f, e$}} {continue}														
							\ForEach{$h$ in \Neighbors{$\{d, e\}$}, $h >a, h \notin \{b,c,f\} $ }{  
								 \lIf {$!$\IsQuadFace{$a, d, h, e$}} {continue}																
								\ForEach{$g \in$ \Neighbors{$\{c,f,h\}$}, $g>a, g \notin \{b,d,e\}$}{  
									\lIf {$!$\IsQuadFace{$d,c,g,h$}} {continue}
									\lIf {$!$\IsQuadFace{$e,f,g,h$}} {continue}
									\lIf {$!$\IsQuadFace{$b,c,g,f$}} {continue}
									\BlankLine
									HexahedronTets $=$ \ComputeTets{$\{abcdefgh\}$}\;
									$H$ $\leftarrow$ $H$ $\cup$ $\{abcdefgh, HexahedronTets\}$\;
								}
							}	
						}					
					}					
				}			
			}
		}
	}	
	\caption{
		Vertex based search algorithm of the set of all potential hexahedra $H$ in a tetrahedral mesh $T$.
	}		
	\label{algo:3D_algorithm}
\end{algorithm}

\subsection{Computing the triangulation of a hexahedron defined by its 8 vertices:}
In addition to the vertices and edges of the hexahedron/prism/pyramid, 
the cell boundary facets and the tetrahedra meshing its interior should be 
computed to fully define the cell.
For the decomposition to be valid, each quadrilateral facet  
should be subdivided into two triangles that are facets of the input tetrahedral mesh.
The existence of these triangle faces must be checked explicitly. 
Indeed, in a 3D triangulation, the existence of edges \{ab\}, \{bc\}, \{bd\}, \{da\} 
does not guarantee that any of the triangles \{abc\}, \{acd\}, \{abd\}, or \{adc\} do exist in the triangulation.
These tests are performed when computing the possible cells  
with Algorithm~\ref{algo:3D_algorithm} in order to skip  invalid configurations and 
accelerate the procedure.
It is then guaranteed, that the boundaries of the cells defined by each set of ordered
vertices output by the combination algorithm correspond to a set of triangular facets of 
the input tetrahedral mesh. 
We also ensure that two merged triangles belong to the same parts of the input model, or model faces.

For each cell, the last step is to determine the interior tetrahedra.
Starting from a tetrahedron that is inside the cell we propagate to the adjacent 
tetrahedra and determine if they are inside the cell too.
For a tetrahedron to be inside the cell, it should either 
(i) have its four vertices be vertices of the cell,
or (ii) have one facet on the boundary and a volume of the same sign than the cell,
or (iii)  be adjacent, through a facet that is not on the theoretical cell boundary, to a tetrahedron that respect (i) or (ii).
The difficulty is that at this step the real cell boundary is not yet determined
since there are two choices to triangulate each quadrilateral facet.
The four triangle facets are then part of what we previously called the theoretical boundary.
All the tetrahedra should belong to the same part of the model, when they do not the cell is discarded.
Note that the boundary tetrahedra, as defined in \S\ref{sec:background}, are
not considered to be inside the cell and are ignored.

\subsection{Efficiency and flexibility of the algorithm}

To improve the  efficiency of Algorithm~\ref{algo:3D_algorithm} or of its prism variation, 
it is crucial to discard invalid or bad quality cells as soon as possible.
The quality and validity of a cell depend only on the coordinates of its vertices.
We recall that we consider that a cell is valid if the Jacobian determinant is
strictly positive at any point of the element.
All cells that have a negative Jacobian determinant are discarded.

The quality of a finite element cell is defined as the minimal value taken by the scaled Jacobian determinant
over the element. If this value is inferior or equal to zero, the cell is invalid.
In the first-order finite element cells we are considering (hexahedra, prisms, and pyramids) 
the maximum quality of the element is bounded by the quality at the corners, itself
bounded by the quality of the facets sharing this vertex $Q_{cell} < min(Q_{corners}) < min(Q_{facets})$.
\begin{itemize}
\item The quality of a quadrilateral face corner $abd$ is evaluated as the sinus of the angle made by the incident edges:
		$$sin( \vec{ab}, \vec{ad})$$.
		
\item The quality of a triangle facet corner $abc$ \cite{johnen_efficient} is evaluated as:
		$$ \frac{2 \ \| \vec{ab}\times \vec{ac}\|}{3\sqrt{3}} \frac{\|\vec{ab}\| + \|\vec{ac}\|+\|\vec{bc}\|  }{\|\vec{ab}\|  \ \|\vec{ac}\| \  \|\vec{bc}\| } $$
		
\item The quality of a hexahedron corner $abde$ is evaluated as the scaled Jacobian:
		$$\frac{ |(\vec{ab} \times \vec{ad})\cdot \vec{ae}|} {\|\vec{ab}\| \ \|\vec{ad}\| \ \|\vec{ae}\|}$$
		
\item The quality of a prism corner $abcd$ is evaluated as: 
	   $$  \frac{2 \ (\vec{ab}\times\vec{ac}) \cdot \vec{ad}}{3\sqrt{3}} \frac{\|\vec{ab}\| + \|\vec{ac}\|+\|\vec{bc}\|  }{\|\vec{ab}\|  \ \|\vec{ac}\| \  \|\vec{bc}\| \ \|\vec{ad}\|}  $$ 
	   
\item The quality of a pyramid base corner $abde$ is evaluated as: 
	$$  \frac{3 \ det(J)^{\frac{2}{3}}}{\|J\|_F ^2 } 
	\ \ \ \ where \ \ \ \ 
		J = J_p J_I^{-1} \ \ \ \ with  \ \ \ \
	 J_I = \begin{pmatrix}
       1 & 0 & \frac{1}{2}       \\[0.3em]
       0 & 1 & \frac{1}{2}       \\[0.3em]
       0 & 0 & \frac{\sqrt{2}}{2} \\
     \end{pmatrix}
	\ \ \ \   J_p = \begin{pmatrix}
       \vec{ab} & \vec{ad} & \vec{ae}       \\[0.3em]
     \end{pmatrix}
	$$
	
    where $\| . \|_F$ denotes the Frobenius norm.	
\end{itemize}

These quality values give an upper bound of the overall cell quality
that we use  to accelerate dramatically the cell identification of Algorithm~\ref{algo:3D_algorithm}.
Indeed a new upper quality bound 
for the cell under construction can be computed when a vertex completing a face or 
a corner is added. When this bound becomes smaller than the required minimum quality, 
the cell construction is terminated.
Additional quality tests on the planarity of quadrilateral facets, or on edge lengths could be added.

To guarantee that the Jacobian determinant is strictly positive,
a lower bound of the quality is needed. That computation is more challenging and time consuming
and is done when the upper bound test passes.
We implemented the mathematically exact tests proposed by \cite{johnen_geometrical_2013,johnen_robust_2017}. 
We additionally modified it so that it is exact and robust to floating point errors.
  
\section{Results}
\label{sec:results}

\subsection{Performances}

\begin{table}[b]
	\centering
	\footnotesize
	\caption{Characteristics of the input tetrahedral meshes on which tests are performed.
	For each model are given the number of vertices, the number of tetrahedra, the number of triangles defining the model boundary,
	the sequential timings to load the mesh and build the data structures used by our algorithm in seconds.
	The meshes are available at: www.hextreme.eu.
	}
	\begin{tabular}{lrrrrr}
		\toprule
		Model                 & \#vertices & \#tets & \#bd tris & Load (s) & Struct. (s) \\
		\midrule
		Cube                  & 127                   & 396                   & 192                   & $<$0.01                  & $<$0.01 \\
		Fusee                 & 11,975                & 50,750                & 15,128                & 0.07                  & 0.04 \\
		CrankShaft            & 23,245                & 104,302               & 27,342                & 0.13                  & 0.09 \\
		Fusee\_1              & 71,947                & 349,893               & 55,954                & 0.40                  & 0.26 \\
		Caliper               & 130,572               & 675,289               & 79,446                & 0.77                  & 0.49 \\
		CrankShaft\_2         & 140,985               & 763,870               & 59,656                & 0.87                  & 0.52 \\
		Fusee\_2              & 161,888               & 828,723               & 98,952                & 0.95                  & 0.58 \\
		FT47\_b               & 221,780               & 1,260,255             & 55,178                & 1.42                  & 0.83 \\
		FT47                  & 370,401               & 2,085,394             & 102,434               & 2.36                  & 1.40 \\
		Fusee\_3              & 501,021               & 2,694,950             & 217,722               & 3.07                  & 1.89 \\
		Los1                  & 583,561               & 3,250,705             & 182,814               & 3.76                  & 2.28 \\
		Knuckle               & 3,058,481             & 17,466,833            & 640,081               & 17.49                 & 16.25 \\
		\bottomrule
	\end{tabular}%
	\label{tab:the_models}%
\end{table}%

We have applied our algorithm to 12 different tetrahedral meshes.
Those were generated using the point placement strategy
described in \cite{baudouin_frontal_2014} and implemented in Gmsh (www.gmsh.info). 
They have between 127 and more than 3 million vertices (Table~\ref{tab:the_models}).

The results of our algorithm in terms of numbers of detected potential hexahedra, prisms, pyramids and 
computational times are given in Table~\ref{tab:our_algo_results}.
The number of potential hexahedra, prisms, pyramids mainly depends on the number of vertices of the input tetrahedral mesh
and on the minimal required quality.
As expected, for a given input mesh, the higher the minimal quality, the faster the algorithm.
For example, the running time on dataset Knuckle decrease from 214s to about 9s 
when quality increases (Table~\ref{tab:our_algo_results}). 
The discard of cells with a too low quality is
key to the efficiency of the algorithm.
The multi-threaded version of our algorithm is very fast with about 300,000 potential hexahedra 
built per second. The algorithm for prisms generates about 900,000 prisms per second 
and the one for pyramids about more than 1.5 millions per second.
All timings and performances are given for a laptop with 
16Go RAM and an Intel$^\circledR$Core\texttrademark i7-6700HQ CPU @2.60 GHz processor.
For all cells, the running time clearly depends 
almost only on the number of potential cells detected.
Note that for models Knuckle and Los1 the computed potential cells are only counted since 
they do not fit in the 16 Go RAM.

We estimate that our method is more than ten times faster than the 
state of the art pattern matching method to identify combinations of tetrahedra into cells 
\citep{sokolov_hexahedral-dominant_2016}.
However, providing a valuable comparison is delicate because quality criteria have a strong impact 
on performances and are not explicitly stated in previous works, preventing comparison.
Moreover no timings are provided for this one step of the hex-dominant meshing workflow.

\begin{table}
   \renewcommand{\arraystretch}{1.}  
   \setlength{\tabcolsep}{3pt} 
  \centering
  \footnotesize
  \caption{
        Number of valid cells (hexahedra, prisms and pyramids) identified in 12 tetrahedral meshes (Table~\ref{tab:the_models}) with our algorithm
		for different minimal quality values.  
        Running times are given for a laptop with 16Go RAM and an Intel$^\circledR$Core\texttrademark  i7-6700HQ CPU @2.60 GHz processor.
	    Input tetrahedral meshes are available at: www.hextreme.eu
	}
  \begin{tabular}{l rrr rrr @{\hspace{3em}} l rrr rrr}
  \toprule
                        & \multicolumn{3}{c}{Cells}                                             & \multicolumn{3}{c}{Timings (s)}                                       &                       & \multicolumn{3}{c}{Cells}                                             & \multicolumn{3}{c}{Timings (s)} \\					
  Q$_{min}$ & \#Hexes& \#Prisms& \#Pyr. & Hex & Pri. & Pyr. & Q$_{min}$ & \#Hexes & \#Prisms & \#Pyr. & Hex & Pri. & Pyr. \\
  \midrule 
  \multicolumn{2}{l}{Cube }                       &                       &                       &                       &                       &                       & \multicolumn{2}{l}{Fusee 2}                       &                       &                       &                       &                       &  \\
  0                     & 710                   & 1,596                 & 1,172                 & 0.03                  & 0.01                  & 0.01                  & 0                     & 4,586,779             & 5,866,663             & 3,122,057             & 9.46                  & 6.88                  & 1.21 \\
  0.2                   & 349                   & 1,052                 & 661                   & $<$0.01                  & $<$0.01                  & $<$0.01                  & 0.2                   & 3,147,148             & 4,614,298             & 2,013,799             & 6.17                  & 5.37                  & 0.92 \\
  0.4                   & 308                   & 1,010                 & 639                   & $<$0.01                  & $<$0.01                  & $<$0.01                  & 0.4                   & 1,940,213             & 3,978,611             & 1,746,731             & 4.13                  & 4.79                  & 0.85 \\
  0.6                   & 129                   & 661                   & 615                   & $<$0.01                  & $<$0.01                  & $<$0.01                  & 0.6                   & 455,155               & 1,849,333             & 1,489,604             & 1.27                  & 2.44                  & 0.78 \\
  0.8                   & 64                    & 218                   & 138                   & $<$0.01                  & $<$0.01                  & $<$0.01                  & 0.8                   & 114,224               & 520,350               & 333,250               & 0.40                  & 0.86                  & 0.46 \\
  \\ 
  \multicolumn{2}{l}{Fusee}&                       &                       &                       &                       &                       & \multicolumn{2}{l}{FT47\_b}                       &                       &                       &                       &                       &  \\ 
  0                     & 149,731               & 251,131               & 158,671               & 0.50                  & 0.37                  & 0.07                  & 0                     & 8,374,128             & 9,845,963             & 4,954,902             & 16.34                 & 11.28                 & 2.06 \\
  0.2                   & 95,259                & 193,440               & 120,496               & 0.30                  & 0.28                  & 0.06                  & 0.2                   & 6,111,946             & 8,046,676             & 3,177,657             & 11.11                 & 9.11                  & 1.45 \\
  0.4                   & 50,931                & 148,390               & 99,949                & 0.19                  & 0.23                  & 0.05                  & 0.4                   & 3,413,741             & 6,963,922             & 2,798,967             & 6.88                  & 8.09                  & 1.34 \\
  0.6                   & 14,978                & 73,397                & 74,831                & 0.07                  & 0.13                  & 0.04                  & 0.6                   & 869,890               & 3,257,991             & 2,349,396             & 2.33                  & 4.13                  & 1.22 \\
  0.8                   & 4,187                 & 20,323                & 18,312                & 0.02                  & 0.05                  & 0.03                  & 0.8                   & 184,921               & 808,736               & 745,059               & 0.65                  & 1.40                  & 0.78 \\
  \\ 
  \multicolumn{2}{l}{CrankShaft}                       &                       &                       &                       &                       &                       & \multicolumn{2}{l}{FT47}                       &                       &                       &                       &                       &  \\ 
  0                     & 309,938               & 516,217               & 327,071               & 1.06                  & 0.78                  & 0.14                  & 0                     & 13,842,934            & 15,994,194            & 8,098,013             & 26.79                 & 19.32                 & 3.42 \\
  0.2                   & 196,441               & 400,698               & 255,842               & 0.65                  & 0.60                  & 0.12                  & 0.2                   & 10,225,710            & 13,177,806            & 4,968,253             & 18.28                 & 14.90                 & 2.33 \\
  0.4                   & 97,194                & 291,381               & 205,388               & 0.38                  & 0.47                  & 0.11                  & 0.4                   & 5,978,752             & 11,638,277            & 4,441,345             & 11.65                 & 13.54                 & 2.18 \\
  0.6                   & 27,549                & 138,308               & 143,964               & 0.15                  & 0.26                  & 0.09                  & 0.6                   & 1,481,030             & 5,392,542             & 3,886,408             & 3.76                  & 6.76                  & 1.99 \\
  0.8                   & 6,381                 & 35,157                & 33,804                & 0.05                  & 0.10                  & 0.06                  & 0.8                   & 325,062               & 1,428,695             & 1,080,358             & 1.05                  & 2.32                  & 1.24 \\
  \\ 
  \multicolumn{2}{l}{Fusee 1}                       &                       &                       &                       &                       &                       & \multicolumn{2}{l}{Fusee 3}                       &                       &                       &                       &                       &  \\ 
  0                     & 1,692,873             & 2,323,521             & 1,283,513             & 3.90                  & 2.85                  & 0.52                  & 0                     & 17,052,534            & 20,251,770            & 10,377,769            & 33.06                 & 24.08                 & 4.43 \\
  0.2                   & 1,098,993             & 1,792,440             & 866,043               & 2.40                  & 2.17                  & 0.40                  & 0.2                   & 12,581,618            & 16,270,311            & 6,218,061             & 22.75                 & 18.75                 & 2.94 \\
  0.4                   & 655,375               & 1,496,380             & 739,789               & 1.58                  & 1.88                  & 0.37                  & 0.4                   & 8,214,177             & 14,639,094            & 5,562,577             & 15.55                 & 16.63                 & 2.75 \\
  0.6                   & 167,752               & 714,225               & 610,352               & 0.53                  & 0.99                  & 0.33                  & 0.6                   & 1,770,306             & 6,701,699             & 4,969,021             & 4.41                  & 8.39                  & 2.59 \\
  0.8                   & 46,215                & 207,852               & 139,521               & 0.17                  & 0.36                  & 0.21                  & 0.8                   & 414,528               & 1,859,542             & 1,047,155             & 1.34                  & 2.92                  & 1.54 \\
  \\ 
  \multicolumn{2}{l}{Caliper}                       &                       &                       &                       &                       &                       & \multicolumn{2}{l}{Los1}                       &                       &                       &                       &                       &  \\ 
  0                     & 3,536,954             & 4,675,695             & 2,508,509             & 7.96                  & 5.73                  & 1.03                  & 0                     & 21,909,206            & 25,424,213            & 12,802,385            & 42.21                 & 30.00                 & 6.30 \\
  0.2                   & 2,341,875             & 3,706,438             & 1,745,660             & 5.07                  & 4.48                  & 0.81                  & 0.2                   & 16,152,752            & 20,505,357            & 7,698,238             & 28.56                 & 23.60                 & 3.61 \\
  0.4                   & 1,262,784             & 3,032,209             & 1,490,139             & 3.15                  & 3.82                  & 0.75                  & 0.4                   & 10,300,168            & 18,629,804            & 6,821,354             & 19.39                 & 21.42                 & 3.37 \\
  0.6                   & 314,028               & 1,372,489             & 1,205,077             & 1.08                  & 1.97                  & 0.67                  & 0.6                   & 2,330,993             & 8,867,751             & 6,106,437             & 5.81                  & 10.84                 & 3.13 \\
  0.8                   & 78,639                & 376,243               & 301,792               & 0.34                  & 0.73                  & 0.43                  & 0.8                   & 502,613               & 2,262,204             & 1,798,365             & 1.61                  & 3.54                  & 1.94 \\
  \\ 
  \multicolumn{2}{l}{Crankshaft 2}                       &                       &                       &                       &                       &                       & \multicolumn{2}{l}{Knuckle}                       &                       &                       &                       &                       &  \\ 
  0                     & 4,406,357             & 5,565,490             & 2,898,887             & 9.24                  & 6.45                  & 1.24                  & 0                     & 86,688,872            & 122,469,704           & 64,495,784            & 220.75                & 161.08                & 28.29 \\
  0.2                   & 2,983,490             & 4,376,413             & 1,874,339             & 5.84                  & 5.11                  & 0.86                  & 0.2                   & 51,459,040            & 93,888,327            & 36,416,856            & 132.79                & 122.33                & 19.81 \\
  0.4                   & 1,683,753             & 3,777,653             & 1,598,601             & 3.63                  & 4.41                  & 0.78                  & 0.4                   & 34,537,750            & 87,271,830            & 34,099,717            & 96.46                 & 115.21                & 19.37 \\
  0.6                   & 417,004               & 1,677,332             & 1,385,062             & 1.20                  & 2.21                  & 0.72                  & 0.6                   & 9,499,091             & 42,600,754            & 32,019,383            & 31.59                 & 62.31                 & 18.77 \\
  0.8                   & 104,728               & 468,154               & 301,721               & 0.36                  & 0.79                  & 0.43                  & 0.8                   & 2,837,726             & 10,931,917            & 6,650,518             & 9.44                  & 20.43                 & 11.32 \\
  \bottomrule
  \end{tabular}%
	\label{tab:our_algo_results}%
\end{table}%

\subsection{Comparison of identified hexahedra with pattern based methods}
\label{sec:comparison}
We compare the number of potential hexahedra identified by our algorithm 
with the number of potential hexahedra that would be identified by 
pattern-matching methods \citep{botella_indirect_2016, sokolov_hexahedral-dominant_2016} 
or vertex combination \citep{yamakawa_fully-automated_2003} 
in Table~\ref{tab:what_we_have_more}.
To count the potential hexahedra matching one of the six cube decompositions or one 
of the four decompositions into seven tetrahedra we compare the dual complex graphs.
To count the potential hexahedra that would be detected by \cite{yamakawa_fully-automated_2003}, we count 
those containing a tetrahedron that has three facets on the hexahedron facets.
On small models, our algorithm detects 4 to 5\% more potential hexahedra than the existing methods.
That number does depend on the input tetrahedral mesh.
On larger meshes, the small difference between methods can be explained by the point placement 
strategy of the input tetrahedral mesh. The points are generated by propagation from the 
boundary of the model. Where the fronts collide, a roughly 2-dimensional surface,
point placement is not optimal. It is in this area that out method makes a difference,
and the bigger the mesh is, the relatively smaller this area.  
Note that the higher the required minimum scaled Jacobian, the smaller the difference between the number of potential
hexahedra detected by our method and the number of hexahedra detected by the existing methods.
This is no surprise since the best quality is obtained for hexahedra that are close to the perfect cube which has
a limited number of decompositions.

\begin{table}
  \centering
  \footnotesize
  \caption{
		Number of valid potential hexahedra detected by our algorithm and 
		comparison with the number of hexahedra that correspond to patterns used by previous 
		methods.
    }
    \begin{tabular}{lrrrr}
\toprule
	Model & Ours & \cite{meshkat_generating_2000} & \cite{botella_indirect_2016, sokolov_hexahedral-dominant_2016} & \cite{yamakawa_fully-automated_2003}\\
\midrule
 Cube                  & 710                   & 672                   & 696                   & 706 \\
  Fusee                 & 149,627               & 123,696               & 142,783               & 146,022 \\
  CrankShaft            & 310,181               & 251,806               & 294,182               & 302,131 \\
  Fusee\_1              & 1,692,188             & 1,532,747             & 1,683,543             & 1,686,709 \\
  Caliper               & 3,536,997             & 3,175,201             & 3,513,863             & 3,522,483 \\
  CrankShaft\_2         & 4,405,892             & 3,919,768             & 4,383,763             & 4,392,081 \\
  Fusee\_2              & 4,585,236             & 4,110,253             & 4,568,332             & 4,574,594 \\
  FT47\_b               & 8,374,930             & 7,384,695             & 8,343,306             & 8,355,231 \\
  FT47                  & 13,846,837            & 12,133,218            & 13,806,781            & 13,821,507 \\
  Fusee\_3              & 17,048,021            & 14,983,008            & 17,010,173            & 17,023,768 \\
  Los1                  & 21,908,307            & 19,308,740            & 21,865,212            & 21,880,467 \\
  Knuckle               & 86,553,836            & 79,544,742            & 86,346,178            & 86,433,942 \\
\bottomrule  		 
	\end{tabular}
  \label{tab:what_we_have_more}%
\end{table}%

\subsection{Hex-dominant meshing}
\label{sec:hex_dominant_meshes}
To demonstrate that hex-dominant meshes may be generated from the set of potential 
cells identified by our algorithm, we choose a subset 
of all compatible hexahedra, prisms, and pyramids.

\paragraph{Cell compatibility}
To have a valid final mesh, chosen cells should be mutually compatible.
Two cells (hexahedron, prism, pyramid) are compatible if all following conditions are satisfied:
\begin{enumerate}
  \item They share no interior tetrahedron.
  \item If they share 4 vertices, they share a quadrilateral face connecting these 4 vertices;
  \item If they share 3 vertices, they share a triangle face connecting these 3 vertices;
  \item If they share 2 vertices, they share an edge connecting these 2 vertices;
\end{enumerate}
The incompatibilities between the remaining tetrahedra (not selected to build any cell)
and the cells should also be accounted for.
There are at least two possible strategies to manage them:
\cite{owen_pyramid_1997} propose to raise pyramids on each non-conformal quadrilateral
face and \cite{botella_indirect_2016} propose to subdivide the 
pyramid or hexahedron incident to a non conformal contact into pyramids and tetrahedra.
Both methods insert a new vertex, the apex of the pyramid or 
a point inside the neighboring cell algorithm. 
Their major drawback is that they increase the proportion of tetrahedra and pyramids 
compared to the proportion of hexahedra. 
In the meshes produced for this paper, the compatibility condition between tetrahedra and 
cells is relaxed. Some quadrilateral faces will be adjacent to one or two triangles.
This mesh should then be used with finite element solvers capable of handling
these type of non-conformities.

\paragraph{Graph formalization}
The selection of the cells of the final mesh can be reformulated as a Maximum Weight Independent 
Set (MWIS) problem \citep{botella_indirect_2016}.
Let us consider the graph $G$ that has one vertex for each of the 
cell that may be built by combining tetrahedra of the input mesh $T$, 
and one edge linking each pair of compatible cells.
The objective is then to find the largest subgraph in which all vertices are linked to one another.
This is the Maximum Clique Problem (MCP), which is in general NP-hard.
We may further associate a weight to each vertex depending on the cell quality.
Since the compatibility graph $G$ is usually very dense, it 
is advantageous to replace it with its complement, the incompatibility graph $G^*$.
The goal is then to find the solution to the Maximum
Weighted Independent Set problem (MWIS).

When the graph $G^*$ contains up to a few hundreds of vertices, the optimal
solution may be found by enumerating all independent sets and comparing
their total weights \cite{wu_review_2015}.
However such an algorithm cannot be expected to terminate in a reasonable amount of
time for graphs with a few thousands of vertices, let alone graphs with a few millions of vertices
like the one we obtain. 
Reviewing the methods to solve the relevant MWIS problem is out of the scope of this paper.
The interested reader is referred to \cite{verhetsel_solving_2017-1} and references 
therein for applicable methods in the specific case of indirect hex-dominant meshing.

\begin{figure}
	\centering
	\includegraphics[width=\textwidth]{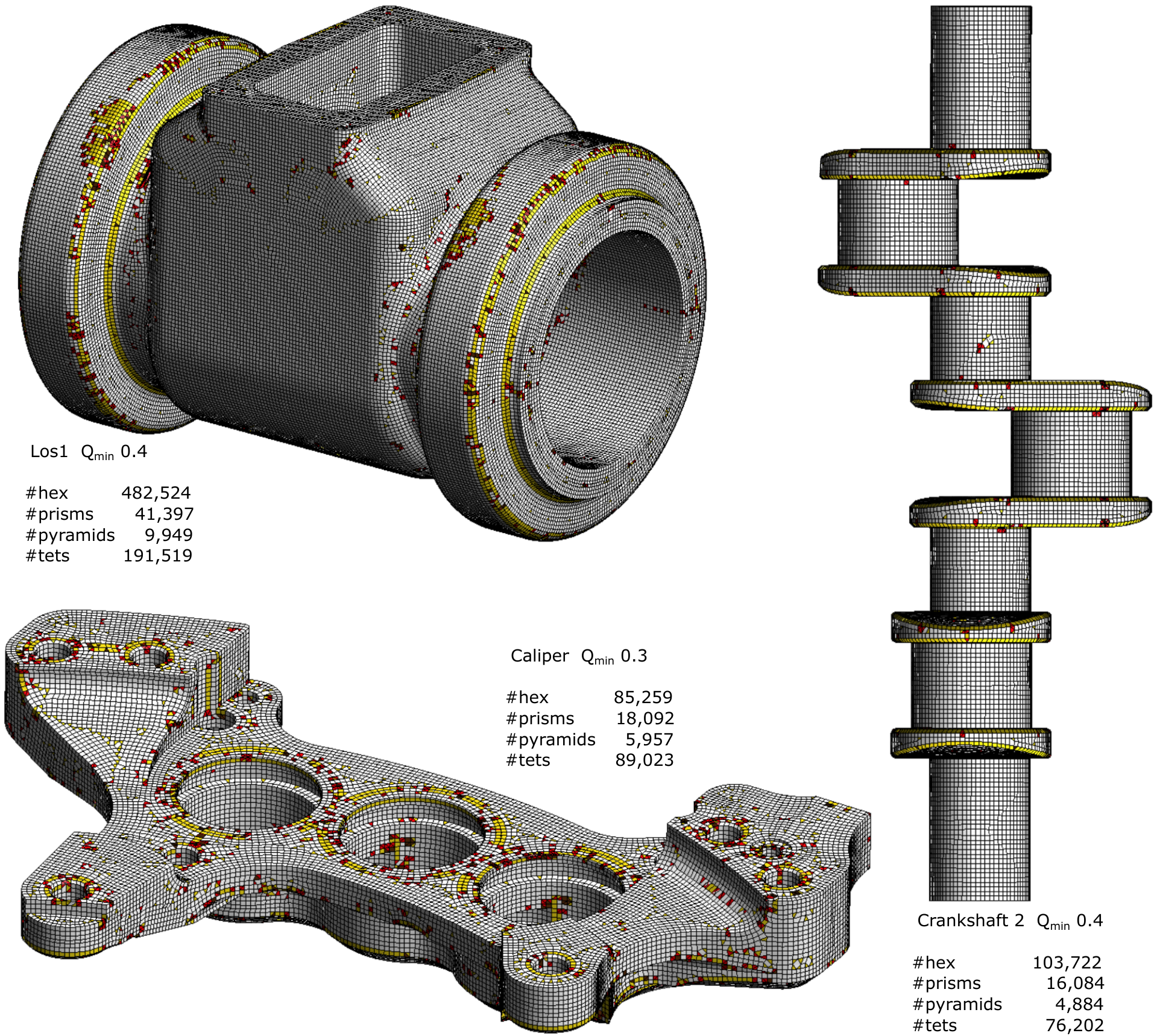}
	\caption{
		Examples of three hexahedral dominant meshes generated by a greedy selection
		(white: hexahedra, red: tetrahedra, yellow: prisms, black: pyramids).
		The complete workflow, from the loading of the tetrahedral mesh till the writing of 
		the mixed-cell mesh, typically runs in less than a minute.
	}
	\label{fig:results}
\end{figure}

\paragraph{Greedy solution}
The strategy we develop to obtain a hex-dominant mesh 
is the one used by previous works:  greedily compute 
an approximate solution to the MWIS problem 
\cite{baudouin_frontal_2014, botella_indirect_2016, sokolov_hexahedral-dominant_2016}. 
The vertices (potential cells) are sorted by decreasing weight (quality), 
and independent vertices (compatible cells) are  iteratively added to the
solution in decreasing order of weight.
This solution could be improved by 
taking advantage of the locality of the problem and 
optimizing small disjoint subgraphs \cite{verhetsel_solving_2017}.
Our non-optimized sequential implementation 
runs typically in a few seconds.
The resulting hex-dominant meshes for three of the input tetrahedral 
meshes are shown on Figure~\ref{fig:results}.

\section{Conclusion}
\label{sec:discussion}
In this paper we solved one step of the indirect hex-dominant meshing workflow that is 
the identification in a given tetrahedral mesh, of all the possible combinations of tetrahedral elements into hexahedra, prisms, pyramids.
Our algorithm identifies all the valid cells elements since no assumption is made on subdivisions into tetrahedra,
each cell  detected only once, bad quality cells are discarded early, and the algorithm is easy to implement.
The C++ code is open-source and available at https://www.hextreme.eu/download/ and 
in Gmsh (www.gmsh.info) \cite{geuzaine_gmsh:_2009}.
Contrary to previous works, our algorithm does not depend on a predefined set of templates,
it is possible to recover hexahedra and prisms that have internal vertices 
and decompositions into unexpected number of tetrahedra.
We have also shown that using a predefined set of templates for the hexahedron 
triangulation does not permit to detect all the potential hexahedra in a given tetrahedral mesh. 
The percentage of the missed potential hexahedra may be significant and reach 5\% when quality requirements are low.

Our algorithm is to be part of a complete workflow to build hex-dominant meshes 
of which all steps have a crucial impact on the final output. 
Indeed, the quality and number of potential hexahedra we can generate highly depend on the input 
tetrahedral mesh and then on the placement of its vertices.
For example is the model Caliper shown on Figure~\ref{fig:results} 
has complex geometrical features and points are non optimally placed around these preventing good quality hex-meshing.
The placement of the vertices is one key to the generation of good hex-dominant mesh
and is a very active research subject.
The choice of the cells of the final mesh is the second key. 
We have shown that there may be up 30 times more hexahedra than they are vertices in the input mesh,
making the construction of the incompatibility graph very costly and the 
exact resolution of the MWIS problem intractable.

\section*{Acknowledgments}
This research is supported by the European Research Council (project HEXTREME, ERC-2015-AdG-694020).


\end{document}